\documentclass{article}
\usepackage{tikz}
\usepackage[compat=1.1.0]{tikz-feynman}
\usepackage[arrowdel]{physics}
\usepackage{amsmath}
\usepackage{graphicx}
\usepackage{listings}
\usepackage{hyperref}
\usepackage[utf8]{inputenc}
\usepackage{caption}
\usepackage{titlesec}
\usepackage{mathtools}
\usepackage[english]{babel}
\usepackage[nottoc]{tocbibind}
\usepackage{setspace} 

\usepackage{vmargin}
\setmarginsrb{2 cm}{1.5 cm}{2 cm}{1.5 cm}{1 cm}{1.5 cm}{1 cm}{1.5 cm}
\usepackage{tabularx}
\usepackage{gensymb}

\titleformat{\paragraph}
{\normalfont\normalsize\bfseries}{\theparagraph}{1em}{}
\titlespacing*{\paragraph}
{0pt}{3.25ex plus 1ex minus .2ex}{1.5ex plus .2ex}

\begin{document}

\title {$g-B_{3}C_{2}N_{3}$: A new potential two dimensional metal-free photocatalyst for overall water splitting}

\author{Sreejani Karmakar,\textit{$^{a}$} Souren Adhikary,\textit{$^{a}$} and Sudipta Dutta\textit{$^{a,b}$}}

\maketitle

\noindent \textit{$^{a}$~Department of Physics and $^{b}$~Center for Atomic, Molecular and Optical Sciences \& Technologies, Indian Institute of Science Education and Research (IISER) Tirupati, Tirupati - 517507, Andhra Pradesh, India}

\begin{abstract}
{In this work, using a hybrid density functional theory (DFT) based calculation, we propose a new two-dimensional (2D) B-C-N material, $g-B_{3}C_{2}N_{3}$, with the promising prospect of metal-free photocatalysis. A comprehensive investigation demonstrates that it is a near ultraviolet (UV) absorbing direct band gap (3.69 eV) semiconductor with robust dynamical and mechanical stability. Estimating the band positions with respect to water oxidation and hydrogen reduction potential levels, we observe that $g-B_{3}C_{2}N_{3}$ monolayer shows the possibility to be used for hydrogen fuel generation through spontaneous solar water splitting, over a broad pH range. Upon biaxial strain application the band gap decreases with increase in tensile strain, leading to a subsequent red shift in absorption spectra, implying enhanced photon harvest under solar irradiation. Furthermore,  due to a combined effect of band gap and work function variation, biaxial strain realigns the band positions, allowing one to control the reducing or oxidizing ability as per requirement to attain environmental sustainability.}
\end{abstract}
\section{Introduction}
Extreme reliance of the global economy on imprudent exploitation of earth's limited resources will lead to a rapid decline in the standard of living very soon \cite{lampert}. For resilience, photocatalysis can be a promising path, addressing some of the major sustainability goals like green energy production, carbon sequestration, water treatment, pollutant degradation and beyond \cite{Barber,Hoffmann,villa}, provided one finds a suitable photocatalyst for that. But designing an efficient and stable photocatalyst with commercial viability has been largely constrained by limited photon harvest, insufficient active sites, high charge recombination and poor charge mobility \cite{Kudo,Ni}. In this regard, the advent of two dimensional (2D) materials, followed by revolutionary progress in nano-engineering has brought a significant surge in the search for appropriate photocatalysts \cite{Ahuja1}. Due to the high surface to volume ratio, 2D materials not only provide a large number of abundant active sites, they offer enhanced exposure for photon harvest and can be used to store energy as well \cite{Xiong,Ahuja2}. Also, the high carrier mobility and reduced migration path length, accompanied by a consequent reduction in recombination rate can significantly improve the photocatalytic performance in reduced dimensions \cite{Li}. Moreover, along with unique physico-chemical properties, 2D materials exhibit easily tunable optoelectronic properties, allowing thorough modulation of photocatalytic activity as per the requirement \cite{Zhen}. Till date many 2D metal oxides ($TiO_{2}$, $WO_{3}$, $ZnO$, $SnO_{2}$)\cite{Khan}, metal chalcogeneides ($MoS_{2}$, $WS_{2}$, $ZnS$, $ZnSe$, $CdS$, $SnS_{2}$)\cite{Zhuang} and some metal-free semiconductors\cite{metal-free1,Son} like phosphorene \cite{Rahman}, carbon nitrides \cite{Cheng} are widely explored as potential photocatalysts, still we are yet to reach the market criteria. For practical application purposes, the metal-free semiconductors have the added advantage of non-toxicity and higher stability against photocorrosion or self-oxidation. So in order to strengthen the photocatalytic technology, finding an efficient metal-free 2D photocatalyst with the ability to perform overall water splitting, is significantly important.

Intensive study on graphene in the context of photocatalysis has claimed it to be an excellent electron acceptor and can be used to improve the performance of composite photocatalysts by enhancing the charge transfer, causing the consequent reduction in charge recombination \cite{graphene}. But to make graphene perform overall water splitting one must introduce a band gap in it, keeping the other advantageous properties, such as ultra high carrier mobility, mechanical and thermal stability intact. Inculcating this idea, it is seen that incorporation of hetero atoms (like oxygen, boron or nitrogen) into the graphene hexagon while maintaining the $sp^{2}$ hybridization can cause the $\pi$-electrons to localize, owing to electronegativity difference between the constituents \cite{Rani,heterodoping}. Hence it turns out to be a powerful strategy to have finite band gap semiconductors for photocatalysis \cite{heterodoping}. For example $BC_{3}$ has a finite gap of 1.83 eV and is predicted to have good photo-oxidizing ability \cite{Omega}. Similarly $C_{3}N$ also has a finite visible-range band gap of 1.05 eV, but that is still less than the net free energy change (1.23 eV) in overall water splitting \cite{Grixti}. Hexagonal boron nitride (h-BN), another graphene analogue in which the honeycomb network is made up of alternating boron and nitrogen, has a wide direct band gap of 5.9 eV \cite{Taniguchi}. Because of deep UV absorbance and consequent poor solar photon harvest, pristine h-BN is unsuitable for photocatalysis. Still keeping in mind the advantages of superior mechanical properties, high chemical and thermal stability, abundance of constituents, significant scientific effort has been put to improve the performance of h-BN based nanomaterials and nanocomposites for environmental and energy applications \cite{Huang}.

Considering the insulating nature of h-BN and semimetallic property of graphene, a systematic amalgamation of both produces a new class of metal-free 2D materials: boron carbon nitride (B-C-N) ternary systems. The possibility of forming planer ring structures with B-C-N is evident from syntheses of graphite like layered $BC_{x}N_{y}$ structures using the methods of chemical vapor deposition (CVD), precursor pyrolysis etc, as can be found in earlier literature starting from the 1970s\cite{Kawaguchi}. However, it was only in the last decade when this effort was translated into successful attempts of synthesizing uniform and continuous 2D hexagonal $B_{x}C_{y}N_{z}$ over a large area \cite{Ajayan1}. These materials generally exhibit a band gap intermediate to that of graphene and h-BN \cite{Angizi}, where the gap-value is dependent on C/BN ratio \cite{Xie}. It has a good carrier mobility similar to graphene \cite{Zhou}, implying a lower possibility of charge recombination. The materials, being made up of boron (B), carbon (C), nitrogen (N), all non-toxic abundant elements, can be scaled to commercial-level cost efficiently. Also, having high mechanical, thermal and chemical stability, these materials will provide long term stability for practical usage\cite{Thomas}. Because of these outstanding properties, in recent years $B_{x}C_{y}N_{z}$ has rightfully grabbed a lot of attention from both academia and industry, owing to its versatile applicability which includes efficient photocatalytic performance as well \cite{Rao}. The experimentally realized hybrid 2D B-C-N systems tend to exhibit prominent phase segregated BN and C domains due to the lattice mismatch (1.5\%) of graphene and h-BN unit cells. Prediction shows that the electronic properties of $B_{x}C_{y}N_{z}$ can be varied to large a extent depending on the composition and configuration of constituting B, C, N elements \cite{Xie,BC2Ntube,BC6N}. So a uniform distribution of B,C and N can unveil new physics and broaden the range of applicability. From first principle calculations it is predicted that such systems, e.g. $BC_{2}N$\cite{BC2N-PC}, $BC_{6}N$\cite{Karmakar} have a promising prospect in photocatalysis. But a uniform distribution of B,C and N demands extremely controlled synthesis conditions and is difficult to achieve experimentally in general. However, in recent experiments homogeneous B(or N) doped and also B-N co-doped nanographene has been synthesized with BCN \cite{BCN-Exp}, $BC_{2}N$ \cite{BC2N-Exp}, $BC_{6}N$\cite{BC6N-Exp} stoichiometry. These experimental progresses have tempted further computational designs of new 2D homogeneous B-C-N materials with broader functionalities and have been the major motivation for the present study on $g-B_{3}C_{2}N_{3}$.

To conduct a systematic investigation of the properties of $g-B_{3}C_{2}N_{3}$, we first perform structural optimization, followed by a thorough stability analysis. Then we explore the electronic and optical properties using hybrid density functional theory. To get a quantitative estimate of catalytic ability, we examine the band edge positions with respect to vacuum and compare it to redox potential levels corresponding to desired reactions. All the results are in favor of the conclusion that $g-B_{3}C_{2}N_{3}$ can serve as a promising photocatalyst under near UV irradiation, over a broad pH range. To improve on the photon harvest efficiency and optimize the band alignment even further, we modulate the opto-electronic properties through biaxial strain application.

\section{Computational details}
All calculations are performed with the help of Vienna Ab Initio Simulations Package (VASP) \cite{vasp1,vasp2}, which implements density functional theory (DFT) using projector augmented-wave (PAW) potentials. A structural optimization is performed first, under generalized gradient approximation (GGA) exchange and correlation functional, as parametrized by Perdew-Burke-Ernzerhof (PBE) \cite{vasp3}. The kinetic energy cut off of the plane wave basis is taken to be 520 eV. The convergence has been achieved with electronic self consistency cut-off of $10^{-6}$ eV and force tolerance of $0.01$ eV/\AA. To sample the Brillouin zone, a $6\times6\times1$ Monkhrost-Pack k-point grid is used. For quantitative description of photocatalytic ability an accurate estimation of band gap is required. That is why Heyd-Scuseria-Ernzerhof (HSE06) hybrid functional with 25\% exact exchange over GGA-PBE is used for all electronic and optical calculations \cite{Heyd}. 

To obtain the absorption spectra, the imaginary part of the dielectric tensor is calculated by summing over the empty states using the equation,

\begin{equation}
       \varepsilon_{\alpha \beta}^{(I)}(\omega)= \frac{4\pi^{2}e^{2}}{\Omega} \lim_{q\to 0} \frac{1}{q^2} \sum_{c,v,\textbf{k}} 2\omega_{\textbf{k}} \delta(\epsilon_{c\textbf{k}}-\epsilon_{v\textbf{k}}-\omega)\times <u_{c\textbf{k}+e_{\alpha q}}|u_{v\textbf{k}}>
       <u_{c\textbf{k}+e_{\beta q}}|u_{v\textbf{k}}>^{*}
 \end{equation}

\noindent where $c$ and $v$ refer to the conduction and valence band, respectively and $u_{c\textbf{k}}$ is the cell periodic part of the orbitals at the wave vector \textbf{k}. Using Kramers-Kronig relation, the real part is obtained as,

\begin{equation}
    \varepsilon_{\alpha \beta}^{(R)}(\omega) = 1 + \frac{2}{\pi} P \int_{0}^{\infty}\frac{\varepsilon_{\alpha \beta}^{(I)}(\omega^\prime)\omega^\prime}{{\omega^\prime}^2 - \omega^2 + i\eta}d\omega'
\end{equation}

\noindent where $P$ indicates the principle value. From these information, the absorption coefficient is calculated as,
\begin{equation}
    \alpha_{\alpha\beta}(\omega) = \frac{\sqrt{2}\omega}{c}[\sqrt{(\varepsilon_{\alpha\beta}^{(I)})^{2}+(\varepsilon_{\alpha\beta}^{(R)})^{2}} - \varepsilon_{\alpha\beta}^{(I)}]^{\frac{1}{2}}
\end{equation}

To compare the position of conduction band minima and valence band maxima with redox potentials, those positions are first calculated by taking vacuum as reference. Then a correspondence with Normal Hydrogen Electrode (NHE) is established by noting that $H^{+}/H_{2}$ reduction level (zero of NHE) is -4.44V at pH 0 with respect to vacuum\cite{Kang}.
For absorption spectra and band edge calculations, VASPKIT is used for post processing \cite{Vaspkit}.

The dynamical matrix is obtained using density functional perturbation theory (DFPT) with single unit cell. A higher energy cut off of 600 eV and a $\Gamma$ centered $12\times12\times1$ k-mesh is used for this. To obtain the phonon frequencies using the DFPT results, Phonopy is used\cite{Togo}.

\section{Results and discussion}
A structural optimization of the $g-B_{3}C_{2}N_{3}$ monolayer system with an eight atom rhombus unit cell (shown in Figure 1a) is performed first. It reveals that $g-B_{3}C_{2}N_{3}$ maintains the hexagonal honeycomb lattice configuration and is fully planar, signifying the characteristics of $sp^{2}$ hybridizaton just like its parent materials. The relaxed lattice vectors, $\vec{a}$ and $\vec{b}$ are of 5.02\AA, making an angle of $60^{\circ}$ with each other. All the structural parameter information along with bond lengths and bond angles are listed in Table.1, which are well in accordance with those of similar materials (e.g. $g-BC_{6}N$, $BC_{3}$, $BC_{2}N$) of this class \cite{Karmakar,Omega,Cohen}.

\begin{figure}[h]
\centering
  \includegraphics[width=1.0\linewidth]{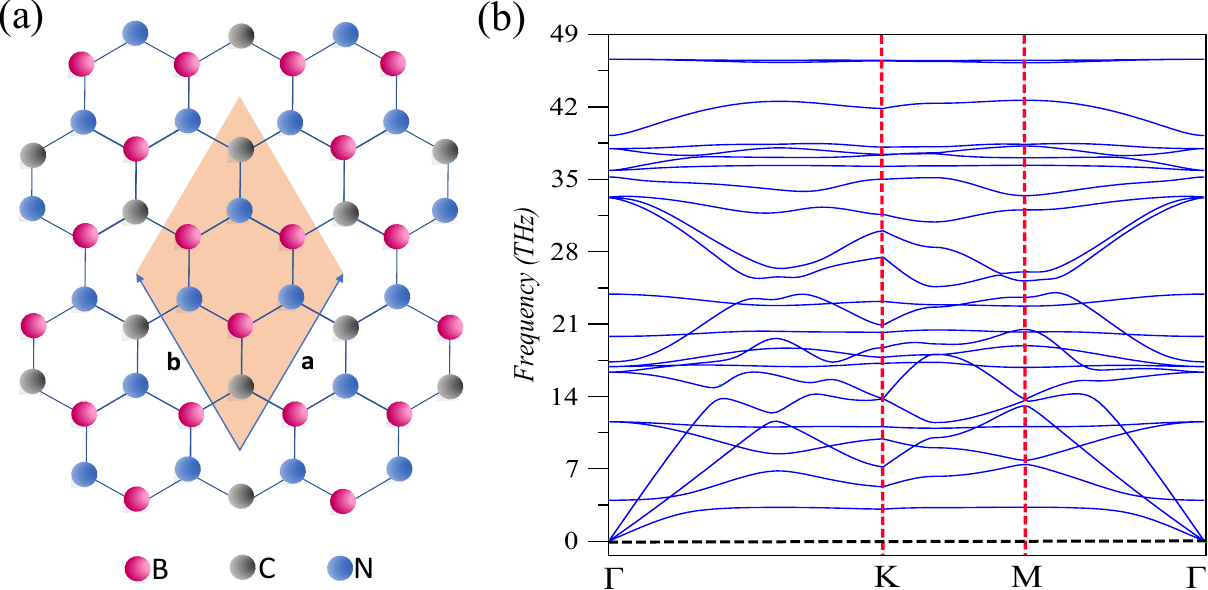}
  \caption{(a) Top view of monolayer $g-B_{3}C_{2}N_{3}$ system. The shaded rhombus indicates the unit cell with lattice vectors, $\vec{a}$ and $\vec{b}$. (b) The phonon spectra of $g-B_{3}C_{2}N_{3}$ system.}
  \label{fgr:example}
\end{figure}

\begin{table}[h]
\small
  \caption{\ The details of the structural configuration in terms of lattice vectors, lattice angle, bond lengths and bond angles.}
  \label{tbl:example1}
  \begin{tabular*}{1\textwidth}{@{\extracolsep{\fill}}llll}
    \hline
    Lattice vectors  & Lattice angle & Bond lengths &  bond angles \\
    (in \AA) & & &  (in \AA) \\
    \hline
    $\vec{a}$ =  5.02 & $\gamma$ = $60^{\circ}$ & C-B = 1.44\AA & $\angle$ NCN = $120^{\circ}$ \\
    $\vec{b}$ =  5.02 &  & C-N = 1.40 \AA & $\angle$ BCB = $120^{\circ}$ \\
     &  & B-N = 1.48 \AA & $\angle$ BNB = $117.42^{\circ}$ \\
     &  &  & $\angle$ NBN = $122.58^{\circ}$ \\
    \hline
  \end{tabular*}
\end{table}

To gain insight on the possibility of experimental realization of $g-B_{3}C_{2}N_{3}$, we conduct a thorough stability analysis. To find out whether the formation of the material is energetically favorable or not, we calculate the formation energy $E_{f}$, using the formula 
\begin{equation}
    E_{f} = E_{B_{3}C_{2}N_{3}} - 3E_{B} - 2E_{C} - 3E_{N}
\end{equation}
where, $E_{B_{3}C_{2}N_{3}}$, $E_{B}$, $E_{C}$, $E_{N}$ are the total ground state energy of relaxed $g-B_{3}C_{2}N_{3}$ unit cell, energy of B atom,  energy of C atom and energy of N atom, respectively. Noting the carbon atom is in the $sp^{2}$ hybridized state, just as in graphene, the term $E_{C}$ is calculated by taking graphene as reference and is given by $E_{CC}/2$, where $E_{CC}$ is the ground state energy of the graphene unit cell. Here one C atom is connected to B atoms, and the other one with N atoms, to emulate a similar atomic environment $E_{B}$ and $E_{N}$ are calculated by taking $BC_{3}$ and $C_{3}N$ system as reference, using the formulae
\begin{equation}
    E_{B} = (E_{BC_{3}} - 3E_{CC})/2
\end{equation}
\begin{equation}
    E_{N} = (E_{C_{3}N} - 3E_{CC})/2
\end{equation}
As per our calculation, $E_{B_{3}C_{2}N_{3}}$ = -68.33 eV, $E_{CC}$ = -18.56 eV, $E_{BC_{3}}$ = -67.71 eV, and $E_{C_{3}N}$ = -70.70 eV, which leads to a $E_{f}$ value of -9.15 eV, evidently showing that formation of $g-B_{3}C_{2}N_{3}$ is energetically favorable.

A free-standing monolayer of $g-B_{3}C_{2}N_{3}$ can exist if and only if it satisfies the dynamical and elastic criteria of stability\cite{stability3}. Dynamical stability check is done by calculating the phonon dispersion relation along the high symmetric path $\Gamma$-K-M-$\Gamma$, as shown in Figure 1b. Absence of any negative frequency value in the phonon dispersion plot implies there is no imaginary eigenvalue corresponding to the dynamical matrix. This result clearly indicates the structural stability of the proposed material. A further calculation of the in-plane stiffness tensor gives an idea about the elastic constants and helps to analyse mechanical stability. Based on symmetry, 2D hexagonal structures have only two independent elastic constants ($C_{ij}$) which are $C_{11}$ and $C_{12}$ whereas $C_{66}$ is given as $C_{66}=(C_{11}-C_{12})/2$ \cite{stability3,Comment,John}. As per our calculation $C_{11}$ and $C_{12}$ are 294.289 N/m and 58.649 N/m, respectively. So the necessary and sufficient criteria of elastic stability for 2D hexagonal materials i.e. $C_{11}>0$, $C_{11}-C_{12}>0$ are both satisfied \cite{Comment,John}. We estimate the Young's modulus of $g-B_{3}C_{2}N_{3}$ to be 282.6 GPa.nm which is in between its parent materials graphene (342 GPa.nm) and h-BN monolayer (276 GPa.nm)\cite{Mapasha}. The Poisson's ratio is 0.199, signifying slightly less brittleness as compared to graphene (0.18) \cite{John}. So the overall analysis leads to the conclusion that the $g-B_{3}C_{2}N_{3}$ demonstrates good mechanical stability under ambient conditions.

The electronic band dispersion (shown in Figure 2a) indicates that $g-B_{3}C_{2}N_{3}$ is a direct band gap semiconductor. The conduction band minima (CBM) and valence band maxima (VBM), both appear at the high-symmetric K point. The electronic band gap value as obtained with GGA-PBE exchange-correlation functional is 2.73 eV. Incorporating 25\% of Hartree-Fock exact exchange along with GGA-PBE, using HSE06 functional, the gap value increases to 3.69 eV. Noting the superiority of HSE06 functional over GGA-PBE, in reproducing experimental reults \cite{Muscat}, we consider 3.69 eV to be the more accurate electronic band gap value. We see that the top most valence band and bottom most conduction band both has contributions from all three constituting elements. However, in the proximity of the direct band gap, the valence band maxima (VBM) and conduction band minima (CBM) are dominantly contributed by N atoms and B atoms, respectively. Unlike graphene, the constituting elements of $g-B_{3}C_{2}N_{3}$, have a finite electronegativity difference, which makes the bonds partially ionic and causes localization of electrons, giving rise to a finite band-gap value. However, the electronegativity difference and hence the degree of localization is much less as compared to the h-BN system. This justifies the band gap value to be intermediate to that of semimetallic graphene and insulating h-BN (5.9 eV)\cite{Taniguchi}, which is consistent with the electronic properties of most of the materials of this B-C-N class\cite{Angizi,BC2N-PC,Karmakar}. 
\begin{figure*}
 \centering
 \includegraphics[width=1.0\linewidth]{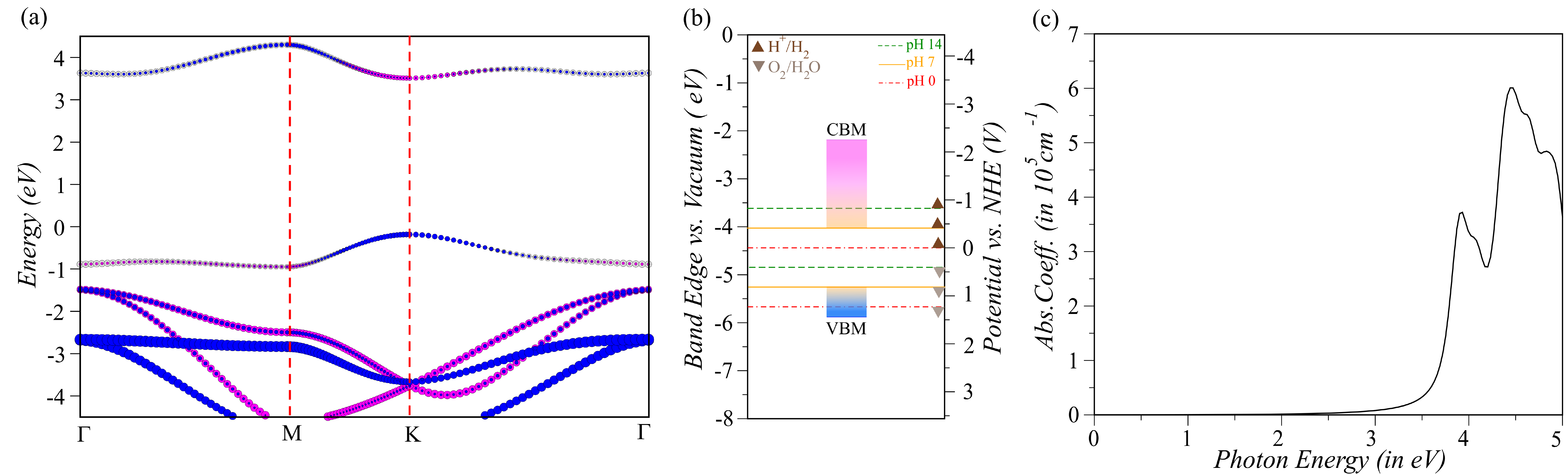}
 \caption{The opto-electronic properties of monolayer $g-B_{3}C_{2}N_{3}$ system as obtained using HSE06 functional. (a)Electronic band dispersion. The pink, grey and blue colors indicate the contributions from boron, carbon and nitrogen, respectively. The radius of the circle is in proportion with the contribution. (b) The position of conduction and valence band edge positions, i.e. band alignment with respect to vacuum (left axis) and Normal Hydrogen Electrode (NHE)(right axis), at different pH values. The pink and blue shaded columns show the reduction and oxidation offsets, respectively for neutral pH. The redox potential levels for pH = 0 and 14 are shown in dotted-dashed and dashed lines, respectively. (c) Absorption spectra, i.e., the absorption coefficient of the system vs. the energy of incoming photon.}
 \label{fgr:example2col}
\end{figure*}

A green photocatalyst must be strong, stable, abundant and non-toxic. Also to drive the catalytic reaction of splitting water into its components (via oxidation) and producing hydrogen fuel out of it (via reduction), the band gap must be greater than the net free energy change (1.23 eV) in the reaction \cite{Walter}. Studying the stability and electronic properties we see that $g-B_{3}C_{2}N_{3}$ satisfies all these requirements. Because of these advantages we explore its photocatalytic ability quantitatively. We observe that the conduction band edge is at a lower potential than the hydrogen reduction potential (-4.44 V) and the valence band edge is at a higher potential than the water oxidation potential (-5.67 V). This means the spontaneous and continuous flow of electrons (or holes) for reduction (or oxidation) is achievable with this material. The reduction offset is 2.25 eV and the oxidation offset is 0.21 eV at pH 0 (shown in Figure 2b). It clearly demonstrates the possibility of hydrogen fuel generation through spontaneous water splitting. To have an idea about the performance in practical scenarios, we varied the pH over the full range and noted its effect on photocatalytic activity. The hydrogen reduction and water oxidation levels at varied pH can be given as,\cite{pH}
\begin{equation}
    E_{H^{+}/H_{2}}^{red} = -4.44 + 0.059\times pH
\end{equation}
\begin{equation}
    E_{O_{2}/H_{2}O}^{ox} = -5.67 + 0.059\times pH
\end{equation}
We find, with pure water (pH 7) the reduction and oxidation offset comes out to be 1.84 eV and 0.62 eV, respectively. The prominent ability to perform simultaneous oxidation and reduction is also maintained even at pH 14, with increased oxidizing ability as compared to neutral pH. So, the variation of pH can be a key to control redox ability. Moreover, in real situations the presence of impurities, sacrificial agents can alter the pH of the water. In that context this robustness of the catalytic ability over the full pH range can be extremely beneficial. 

To have an idea about the energy of the harvested photon to be used in photocatalysis, the absorption spectra with HSE06 functional is calculated (shown in Figure 2c). It reveals that the material harvests photons in near UV range, in contrast to the far UV absorbance of h-BN \cite{Taniguchi}. The plot indicates the appearance of absorption edge at 3.69 eV. It establishes that the optical gap is the same as the electronic gap (3.69 eV). This fact can be attributed to the direct nature of the band gap. The allowed direct optical transitions make this material advantageous for optical device designing in general. But for photocatalysis even though it has a positive influence due to high photon harvest, it may also affect the process adversely by increasing the recombination of free carriers. While engineering practical devices, this needs to be taken care of separately in some other way like facet engineering, co-catalyst loading etc \cite{Takata,Qin}.
\begin{figure}[h]
    \centering
    \includegraphics[width=1.0\linewidth]{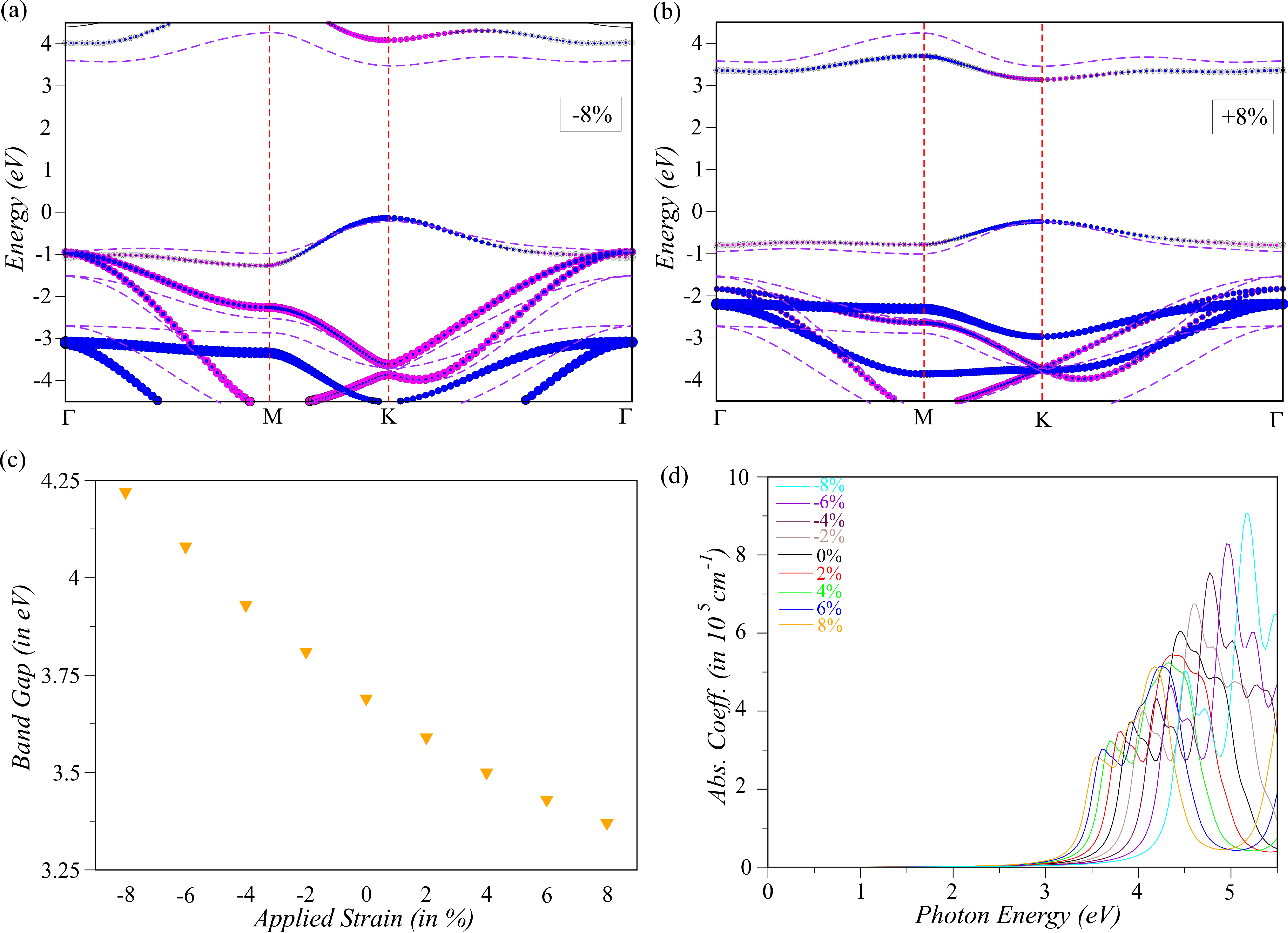}
    \caption{ The variation in opto-electronic properties of monolayer $g-B_{3}C_{2}N_{3}$ system due to applied biaxial strain, as obtained using HSE06 functional. Electronic band dispersion for system with (a) 8\% compression and (b) 8\% expansion. The pink, grey and blue colors indicate the contributions from boron, carbon and nitrogen, respectively. The radius of the circle is in proportion with the contribution. The dashed lines represent the electronic dispersion for unstrained system for comparison. (c) The variation in band gap value in the range of -8\% to +8\% biaxial strain. (d) The variation in absorption spectra in the range of -8\% to +8\% biaxial strain. }
 \label{fig:my_label}
\end{figure}

In spite of having the prominent ability to perform over all water splitting, the near UV absorbance of $g-B_{3}C_{2}N_{3}$ can limit the overall photoconversion efficiency while performing photocatalysis using solar photons. Since solar spectra peaks at visible range, the number of harvested photons under solar irradiation itself will be low in this range. So to tune the photocatalytic efficiency of $g-B_{3}C_{2}N_{3}$ further, we apply biaxial strain in the range of -8\% to +8\%. The strain quantifying parameter is defined as $\delta = (a_{s}-a_{0})/a_{0}$, where $a_{0}$ and $a_{s}$ are the lattice constants at relaxed and strained condition, respectively. So a positive value of $\delta$ signifies expansion ($a_{s}$ $>$ $a_{0}$) and negative value indicates compression ($a_{s}$ $<$ $a_{0}$). Experimentally one can achieve this by introducing lattice mismatch or by applying external load \cite{Dai,JLiu}. 

Studying the variation of electronic dispersion with respect to applied biaxial strain, we observe that with compression, the bands become more dispersive near extrema points and the band gap gradually widens (shown in Figure 3a). This can be attributed to the increased coupling strength of orbitals due to decreased bond length during compression, as is seen in the literature\cite{Kuo}. On the contrary, expansion leads to less dispersive bands with reduced gap value (shown in Figure 3b), because of the decreased coupling strength. In our previous work on $BC_{6}N$, a similar trend was observed \cite{Karmakar}. We observe that the band gap of $g-B_{3}C_{2}N_{3}$ can be tuned over a wide range, from 4.22 eV to 3.37 eV (shown in Figure 3c), upon biaxial strain application. The reduction of the band gap under tensile strain is manifested in a subsequent red shift in the absorption spectra as shown in Figure 3d. This will ensure higher photon harvest under solar irradiation and will increase the efficiency of the entire process. 

\begin{figure}[h]
    \centering
    \includegraphics[width=1.0\linewidth]{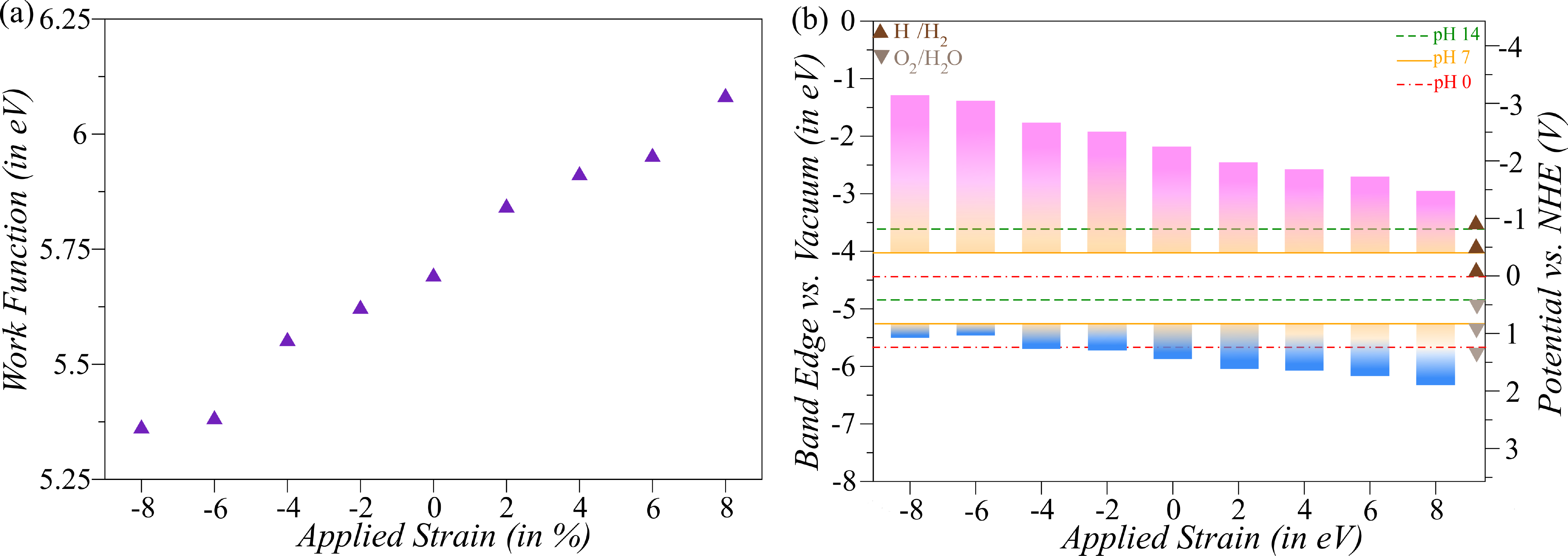}
    \caption{ (a) The variation in work function of monolayer $g-B_{3}C_{2}N_{3}$ system due to applied biaxial strain in the range of -8\% to +8\%, as obtained using HSE06 functional. (b)The band alignment for monolayer $g-B_{3}C_{2}N_{3}$ at different strained condition (-8\% to +8\%) with respect to vacuum (left axis) and Normal Hydrogen Electrode (NHE)(right axis), at different pH values. The pink and blue shaded columns show the reduction and oxidation offsets, respectively for neutral pH. The redox potential levels for pH = 0 and 14 are shown in dotted-dashed and dashed lines, respectively. }
 \label{fig:my_label}
\end{figure}

The work function shows a trend opposite to that of the band gap. It decreases with compression and increases with expansion (shown in Figure 4a). The same trend is reported earlier for graphene, h-BN and BCN systems \cite{strain}. This can be explained from the band dispersion variation itself. Higher dispersive bands in the compression region clearly signifies lower effective mass and higher mobility as compared to the expanded scenario with less dispersive bands. Since the electrons are comparatively less localized in the compressed system, the minimum energy required for the electrons to be freed from the electrostatic potential, i.e. the work function will be low as compared to that for the expanded case. This means the position of Fermi energy shifts to higher potential level with increase in strain. As a combined effect of band gap variation and shift in Fermi energy position, the reducing ability increases with compression and decreases with expansion. Whereas the oxidizing ability follows exactly the opposite trend. The variation of redox offsets due to applied biaxial strain at different pH values are shown in Figure 4b. For compression higher than 6\%, the material completely loses its oxidizing ability at highly acidic pH. Even though in the mentioned scenario it can not perform overall water splitting, it can still act as a cathode in photoelectrocatalysis. In all other cases, the ability to perform spontaneous overall water splitting is maintained throughout the entire pH range. In case of expansion at neutral pH, the oxidizing and reducing offsets are comparable, which brings a better balance in the rates of oxygen evolution reaction (OER) and hydrogen evolution reaction (HER). This will aid the cyclic charge flow during the catalytic reactions and improve the usage of free carriers, boosting the efficiency of the process. As a combined effect of pH and strain variation one can get dominance of reducing ability (at acidic pH, under compression) or oxidizing ability (at basic pH, under expansion) as per requirement. This will give more control over selectivity of the photocatalytic reactions depending on the application.

\section{Conclusions}
The present study establishes that $g-B_{3}C_{2}N_{3}$ has prominent redox offsets and can act as a green hydrogen fuel producing material, over a broad pH range. This ability of performing simultaneous reduction and oxidation can also be exploited for bio-fuel generation, heavy metal ion deposition for water purification, oxidative decomposition organic pollutants etc, depending on which catalytic reactions are being performed. Changing the condition of pH and/or strain, $g-B_{3}C_{2}N_{3}$ can be made to perform only reduction or oxidation or both, that too with variable offset. This allows one to control the selectivity of photocatalytic reactions or suppress side reactions as per requirement. Moreover, for over all water splitting, tensile strain provides the added advantage of red shifted absorption spectra and balanced HER-OER offsets. As a whole we see that biaxial strain has significant positive impact in the photocatalytic ability. With these results we claim $g-B_{3}C_{2}N_{3}$ to be a promising potential photocatalyst and can contribute in paving the way towards environmental sustainability.

\section*{Acknowledgements}
The authors thank IISER Tirupati for Intramural Funding and SERB, Dept. of Science and Technology (DST), Govt. of India for research grant CRG/2021/001731. The authors acknowledge National Supercomputing Mission (NSM) for providing computing resources of ‘PARAM Brahma’ at IISER Pune, which is implemented by C-DAC and supported by the Ministry of Electronics and Information Technology (MeitY) and DST, Govt. of India.

\clearpage

\end{document}